\documentclass{bmcart}

\usepackage[utf8]{inputenc} %unicode support
\usepackage{graphicx}
\usepackage{enumitem}
\usepackage[colorinlistoftodos]{todonotes}
\usepackage{pdfpages}

\startlocaldefs
\endlocaldefs

\begin{document}

%%% Start of article front matter
\begin{frontmatter}

\begin{fmbox}
\dochead{Research}


\title{Mechanism, measurement, and quantification of stress in decision process: a model based systematic-review protocol}

%%%%%%%%%%%%%%%%%%%%%%%%%%%%%%%%%%%%%%%%%%%%%%
%%                                          %%
%% Enter the authors here                   %%
%%                                          %%
%% Specify information, if available,       %%
%% in the form:                             %%
%%   <key>={<id1>,<id2>}                    %%
%%   <key>=                                 %%
%% Comment or delete the keys which are     %%
%% not used. Repeat \author command as much %%
%% as required.                             %%
%%                                          %%
%%%%%%%%%%%%%%%%%%%%%%%%%%%%%%%%%%%%%%%%%%%%%%

\author[
   addressref={aff1},                   % id's of addresses, e.g. {aff1,aff2}
                                         % id of corresponding address, if any
%   noteref={n1},                        % id's of article notes, if any
   %email={jane.e.doe@cambridge.co.uk}   % email address
]{\inits{CS}\fnm{Chang} \snm{Su}}
\author[
   addressref={aff2},
]{\inits{XL}\fnm{Xiaoyuan} \snm{Li}}
\author[
   addressref={aff3,aff4},
]{\inits{LY}\fnm{Lin} \snm{Yang}}
\author[
   addressref={aff1},
    corref={aff1},  
   email={yong.zeng@concordia.ca}
]{\inits{YZ}\fnm{Yong } \snm{Zeng}}

%%%%%%%%%%%%%%%%%%%%%%%%%%%%%%%%%%%%%%%%%%%%%%
%%                                          %%
%% Enter the authors' addresses here        %%
%%                                          %%
%% Repeat \address commands as much as      %%
%% required.                                %%
%%                                          %%
%%%%%%%%%%%%%%%%%%%%%%%%%%%%%%%%%%%%%%%%%%%%%%

\address[id=aff1]{%                           % unique id
  \orgname{Concordia Institute for Information Systems Engineering, Concordia University}, % university, etc
  %\street{Waterloo Road},                     %
  \postcode{H3G 1M8}                                % post or zip code
  \city{Montreal},                              % city
  \cny{ Canada}                                    % country
}

\address[id=aff2]{%
  \orgname{ School of Electrical Engineering, Zhengzhou University},
  %\street{D\"{u}sternbrooker Weg 20},
  %\postcode{24105}
  \city{Zhengzhou},
  \cny{China}
}

\address[id=aff3]{%
  \orgname{Department of Cancer Epidemiology and Prevention Research, Alberta Health Service},
  %\street{D\"{u}sternbrooker Weg 20},
  %\postcode{24105}
  \city{Calgary},
  \cny{Canada}
}

\address[id=aff4]{%
  \orgname{ Departments of Oncology and Community Health Sciences, University of Calgary},
  %\street{D\"{u}sternbrooker Weg 20},
  %\postcode{24105}
  \city{Calgary},
  \cny{Canada}
}

%%%%%%%%%%%%%%%%%%%%%%%%%%%%%%%%%%%%%%%%%%%%%%
%%                                          %%
%% Enter short notes here                   %%
%%                                          %%
%% Short notes will be after addresses      %%
%% on first page.                           %%
%%                                          %%
%%%%%%%%%%%%%%%%%%%%%%%%%%%%%%%%%%%%%%%%%%%%%%

\begin{artnotes}
%\note{Sample of title note}     % note to the article
%\note[id=n1]{Equal contributor} % note, connected to author
\end{artnotes}

\end{fmbox}% comment this for two column layout

%%%%%%%%%%%%%%%%%%%%%%%%%%%%%%%%%%%%%%%%%%%%%%
%%                                          %%
%% The Abstract begins here                 %%
%%                                          %%
%% Please refer to the Instructions for     %%
%% authors on http://www.biomedcentral.com  %%
%% and include the section headings         %%
%% accordingly for your article type.       %%
%%                                          %%
%%%%%%%%%%%%%%%%%%%%%%%%%%%%%%%%%%%%%%%%%%%%%%

\begin{abstractbox}

\begin{abstract} % abstract
\parttitle{Background:} %if any
Every human action begins with decision-making. Stress is a significant source of biases that can influence human decision-making. In order to understand the relationship between stress and decision-making, stress quantification is fundamental. Different methods of measuring and quantifying stress in decision-making have been described in the literature while an up-to-date systematic review of the existing methods is lacking. Moreover, mental stress, mental effort, cognitive workload, and workload are often used interchangeably but should be distinguished to enable in-depth investigations of decision-mechanisms. Our objectives are to clarify stress related concepts and review the measurement, quantification, and application of stress during decision making activities.

\parttitle{ Methods:} %if any
We developed and followed a systematic reviews protocol to analyze the literature related to stress in decision-making. We systematically searched Web of Science, Scopus, PubMed, and ERIC (EBSCO) between 1990 and 2020 with English language. We will include any literature reporting measurable stress-related outcomes, including stress, workload, cognitive workload, mental effort during the decision-making process. All research designs investigating the quantification and measurement of stress for healthy adults are eligible for this review. Research postulates are proposed based on the stress-effort model. Two reviewers will independently screen the articles for inclusion and assess for study quality  using the Quality Assessment of Diagnostic Accuracy Studies-2 tool. We will extract data from each study including research objective, research method, research domain, triggering method, objective measure, subjective measure, and research results for knowledge syntheses.

\parttitle{Discussion:}
 The physiological responses to stress are proposed for verification. This systematic review will provide knowledge on decision mechanisms under stress and stress related concepts. The improved understanding on stress measurement, quantification, and application in specific research context will advance methods to study and optimize decision making.

\end{abstract}

%%%%%%%%%%%%%%%%%%%%%%%%%%%%%%%%%%%%%%%%%%%%%%
%%                                          %%
%% The keywords begin here                  %%
%%                                          %%
%% Put each keyword in separate \kwd{}.     %%
%%                                          %%
%%%%%%%%%%%%%%%%%%%%%%%%%%%%%%%%%%%%%%%%%%%%%%

\begin{keyword}
\kwd{ systematic review}
\kwd{protocol, stress}
\kwd{acute stress}
\kwd{workload}
\kwd{cognitive workload}
\kwd{mental effort}
\kwd{stress measures}
\kwd{stress quantification}
\end{keyword}

% MSC classifications codes, if any
%\begin{keyword}[class=AMS]
%\kwd[Primary ]{}
%\kwd{}
%\kwd[; secondary ]{}
%\end{keyword}

\end{abstractbox}
%
%\end{fmbox}% uncomment this for twcolumn layout

\end{frontmatter}

%%%%%%%%%%%%%%%%%%%%%%%%%%%%%%%%%%%%%%%%%%%%%%
%%                                          %%
%% The Main Body begins here                %%
%%                                          %%
%% Please refer to the instructions for     %%
%% authors on:                              %%
%% http://www.biomedcentral.com/info/authors%%
%% and include the section headings         %%
%% accordingly for your article type.       %%
%%                                          %%
%% See the Results and Discussion section   %%
%% for details on how to create sub-sections%%
%%                                          %%
%% use \cite{...} to cite references        %%
%%  \cite{koon} and                         %%
%%  \cite{oreg,khar,zvai,xjon,schn,pond}    %%
%%  \nocite{smith,marg,hunn,advi,koha,mouse}%%
%%                                          %%
%%%%%%%%%%%%%%%%%%%%%%%%%%%%%%%%%%%%%%%%%%%%%%

%%%%%%%%%%%%%%%%%%%%%%%%% start of article main body
% <put your article body there>

%%%%%%%%%%%%%%%%
%% Background %%
%%
\section*{Background}

Every human action begins with decision-making, which is an established discipline that identifies and selects alternatives based on the values and preferences of decision-makers \cite{fulop2005introduction}. Decision-making can be characterized as the organism's ability to choose the most adaptive action and outcomes from all possible alternatives \cite{bechara2000emotion}.

Stress is a significant source of biases and can influence human decision-making \cite{starcke2012decision,pabst2013stress,keinan1987decision}. Stress can be defined as 'the nonspecific result of any demand upon the body' \cite{selye1976stress}. This definition indicates that stressor, as the load or stimulus, triggers stress response \cite{Selye_1950}. A stressor can be a physical trigger that threatens homeostasis \cite{chrousos2013mechanisms,chrousos1992concepts,chrousos2009stress}, such as injury, noise, and extreme temperature. A stressor can also be a psychological trigger, such as time-pressured tasks, trauma events \cite{resick2014stress}, and unexpected events. The stress response represents the body's attempt to counteract the stressor and re-establish homeostasis \cite{chrousos2013mechanisms}, such as increased blood pressure, the increased cortisol levels, and the decreased heart rate variability. Hence, these physiological responses can be proxy measures for stress. The entire process, from the stimulus activation to the individual response, is a decision-making process. Stress can bias decision-making strategies, affecting the individual decision-making ability to perform actions, and may cause mistakes and incorrect decisions \cite{saleem2011effect,dias2009chronic}.

\begin{figure}[h!]
  \centerline{\includegraphics{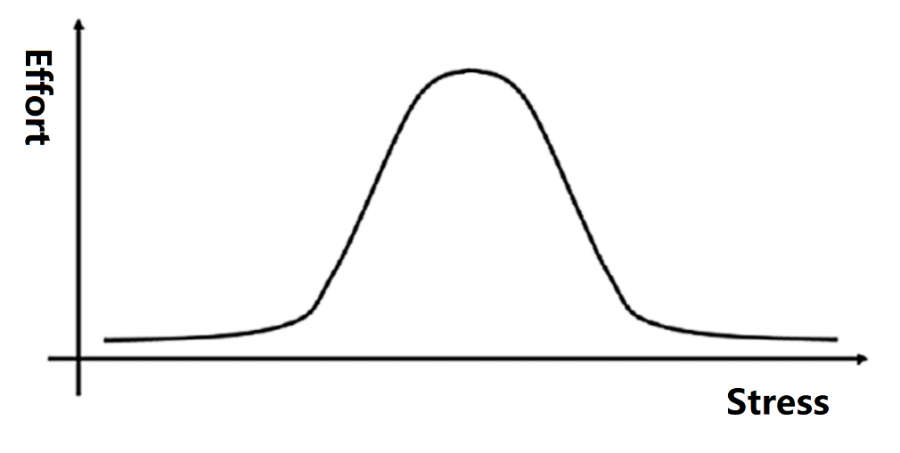}}
  \caption{\label{fig:fg1}
  \csentence{The stress-effort model.}
      Source: Nguyen and Zeng \cite{nguyen2014physiological}}
      \end{figure}

Different levels of stress have different impacts on decision-making. Stress can be classified into "eustress" (positive stress) and "distress" (negative stress) \cite{lazarus1984stress}. Eustress is considered as constructive stress because it will not cause any issues in solving and pursuing goals. In contrast, it can stimulate the individual to feel happy or motivated, which improves productivity and performance. On the contrary, distress is considered as destructive stress because it is more challenging than eustress to deal with and can become a source of barriers to achieving goals \cite{yang2021implementation} . Distress can damage cognitive abilities and reduce individual performance. The Yerkes-Dodson law illustrates the relationship between stress and performance in an inverse U-shaped curve \cite{yerkes1908relation}. However, the Yerkes-Dodson law did not specify factors that influence stress. Furthermore, for simple problems, people can perform excellently with minimal stress. Nguyen and Zeng thus revised the Yerkes-Dodson law and proposed the stress-effort model \cite{nguyen2014physiological}, as shown in Figure \ref{fig:fg1}. Given that both the presence and the level of stress could influence decision-making, it is critical to measure and quantify stress.

Likewise, the stress-effort model shows a bell-shaped curve relationship between stress and effort. The effort may enhance with stress, up to a point, then it will decline with continuously increased stress. Nguyen and Zeng also developed a conceptual stress equation to describe the relationship of stress with the perceived workload, knowledge, skills, and affect \cite{nguyen2012theoretical}, as shown in a mathematical form below:
\begin{eqnarray}\label{eq1}
\sigma = \frac{w_{p}}{(k+S)*\alpha}
\end{eqnarray}
where $\sigma$ represents stress, and $W_{p}$,$K$,$S$, and $\alpha$ represents perceived workload, knowledge, skills, and affect, respectively. 

To measure stress is to measure the stress perceived by individuals \cite{fry1958experiments}. Researchers measures the difference from stressful situation to the normal situation (baseline) of individuals. The process of stress activation and the process of stress recovery can make this difference. Moreover, how quickly people recover from the episode of stress can determine whether it is acute stress (short-term stress) or chronic stress (long-term stress) \cite{mcgonagle1990chronic,katz1981acute}. Acute stress will develop rapidly and last for a short time through a particularly stressful event. Chronic stress can bring depression, personality disorders, and diseases \cite{maeng2017post,checkley1996neuroendocrinology,vitaliano2002path}. Our review will focus on acute stress, which has been the central focus of decision-making \cite{wemm2017effects}.

It can be seen from Equation \ref{eq1} that stress is related to workload, knowledge, emotion, perception, cognitive and affective capabilities. Previous research on the relationship between stress and decision-making had involved concepts, such as mental effort, workload, cognitive workload \cite{alsuraykh2019stress,matthews2017metrics,shenhav2017toward,magnusdottir2017cognitive,recarte2003mental,fothergill2008effect}. These stress-related concepts sometimes are used interchangeably to describe the same phenomenon \cite{lundberg1999stress,hockey1997compensatory} and sometimes the same concept had been used to refer to different phenomena \cite{ahlstrom2006using,marshall2002index}. Correctly distinguishing these concepts can improve the interpretations of findings from stress-related experiments or applications. Therefore, this review will integrate these stress-related concepts to elucidate accurate stress measurement and quantification through a decision-making mechanism.

Stress measures can be categorized into subjective and objective measures. Subjective measures assess stress using self-report protocols or questionnaires \cite{fulop2005introduction, bechara2000emotion,starcke2012decision, pabst2013stress,cohen1983global,hart2006nasa}. In contrast, objective measures use devices to assess physiological, physical, or behavioural responses \cite{hellhammer2009salivary,alba2019relationship,zhao2020teeg,hjortskov2004effect}. Both measures have their unique advantages and disadvantages. Subjective measures are perceived to have high reliability and user trust \cite{goyal2016automation}. However, as self-report are retrospective measures, subjective measures can induce recall bias \cite{fry1958experiments}, which may cause errors and biases that could compromise their validity. Meanwhile, objective measures could have high validity \cite{goyal2016automation}. Taking advantage of the rapid development of sensor technologies and data science, objective measures have become increasingly popular in recent years \cite{nguyen2014physiological,hellhammer2009salivary,tang2009quantifying,hosseini2010emotional}. While the adoption of objective measures brings significant benefits to stress measurement and quantification, advanced interdisciplinary knowledge is critical to their successful applications. Furthermore, both subjective and objective measures are combined to achieve a comprehensive stress assessment \cite{nguyen2014physiological,al2015mental}. Therefore, this review will summarize the experiment protocols of stress measures, categorizes measures according to their functions and applications, and survey and assess the stress quantification algorithms and processing methods. 

\section*{Methods}

\subsection*{Research postulates}

\begin{figure}[h!]
  \centerline{\includegraphics[width=1.0\textwidth]{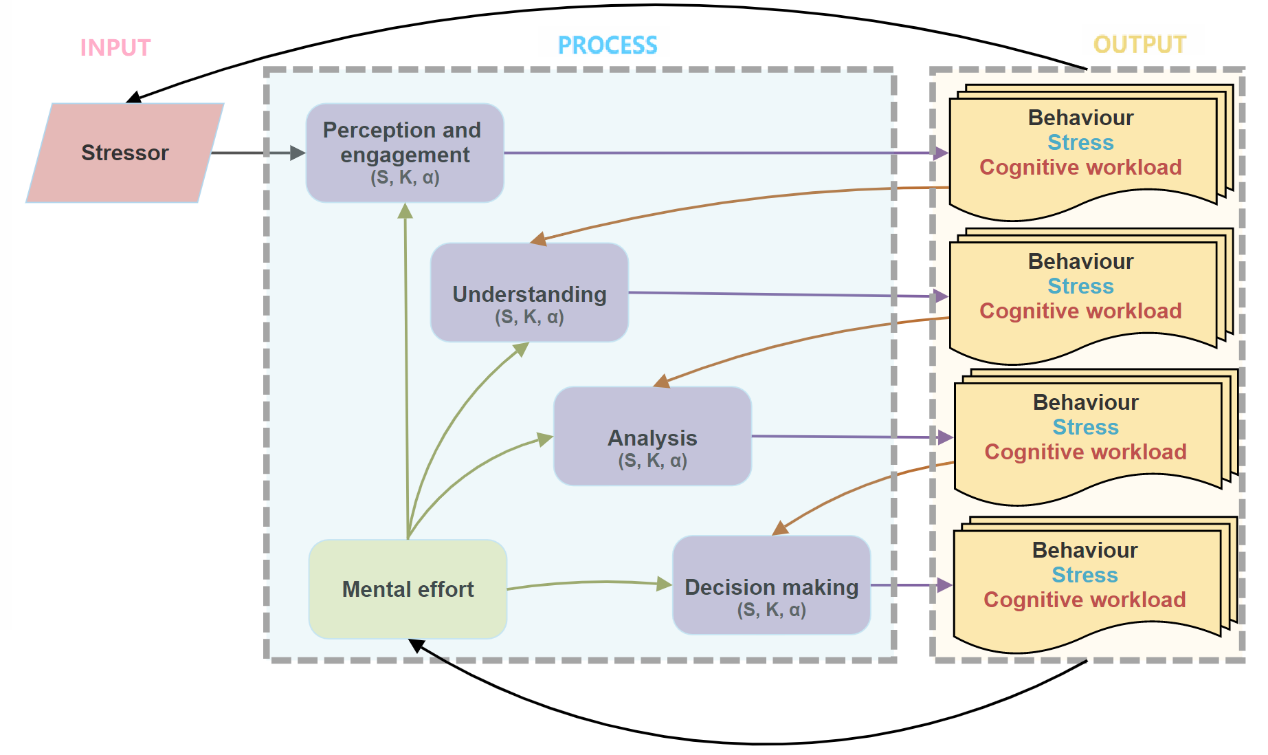}}
  \caption{\csentence{\label{fig:fg2}General decision mechanism.}
       $K$ represents knowledge, $S$ represents skills, and $\alpha$ represents affect.}
      \end{figure}

The foundation of this review comes from the stress-effort model shown in Figure \ref{fig:fg1} \cite{nguyen2014physiological}. Herein, we propose a general decision mechanism,  (as shown in Figure \ref{fig:fg2}), which to integrates stress and stress-related concepts into the stress-effort model.

The general decision mechanism consists of three components specified below: 
\begin{itemize}
\item Input: The input of the decision mechanism is a stressor. 
\item Process: The decision-making process involves four distinctive stages, including high-level cognitive activities \cite{cromley2010cognitive} such as: “perception and engagement”, “understanding”, “analysis”, and “decision making”. 
\item Output: The output consists of everything produced by the decision-making process, such as behaviours, stress, and cognitive workload \cite{bodala2014cognitive}.
\end{itemize}

A stressor, as the input, stimulates an individual to start the decision-making process. Guided by the individual’s knowledge, skills, and affect, each stage of the decision-making process leads to certain behaviours while taking corresponding mental effort and triggering consequential mental stress \cite{bodala2014cognitive}. In Figure \ref{fig:fg2}, behaviours contains physical actions associated with individual psychomotor skills and abilities \cite{chaiken2000organization}. In addition, a cognitive workload will be created and becomes the input for the next stage. As the process continues to evolve, cognitive workload, stress, and behaviour are recursively updated.

\begin{figure}[h!]
\centerline{\includegraphics[width=1.0\textwidth]{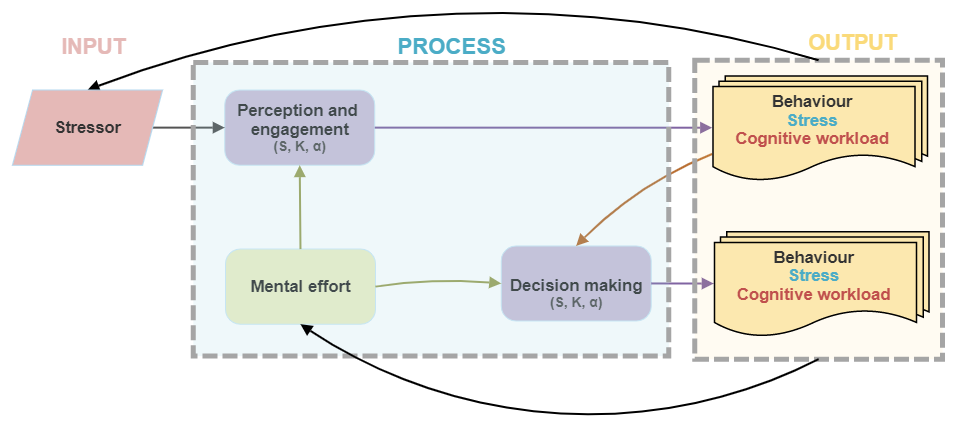}}
  \caption{\label{fig:fg3}
  \csentence{Automatic decision mechanism under extreme stresses}
    }
      \end{figure}

While decision-making can be a controlled, rational, and effortful process, it can also be automatic, unconscious, emotional, and effortless. During the rational decision-making process, people can reason with their knowledge and skills to deal with their cognitive workload. In contrast, automatic decision-making has no self-awareness or control of the mind \cite{daniel2017thinking}. The former occurs when the cognitive workload is mild, whereas the latter kicks in when the stress levels are excessively heavy or lacking  \cite{nguyen2012theoretical}. Figure \ref{fig:fg2} describes the rational decision-making process, and Figure \ref{fig:fg3} illustrates the emotional (automatic) process of decision-making. 

Like the rational decision-making mechanism shown in Figure \ref{fig:fg2}, the automatic decision-making mechanism also includes three components: input, process, and output but with only two stages, which are “perception and engagement” and “decision making”. Thus, people would make their decisions based only on perception and engagement while skipping the understanding and analysis stages. 

\begin{figure}[h!]
\centerline{\includegraphics[width=1.0\textwidth]{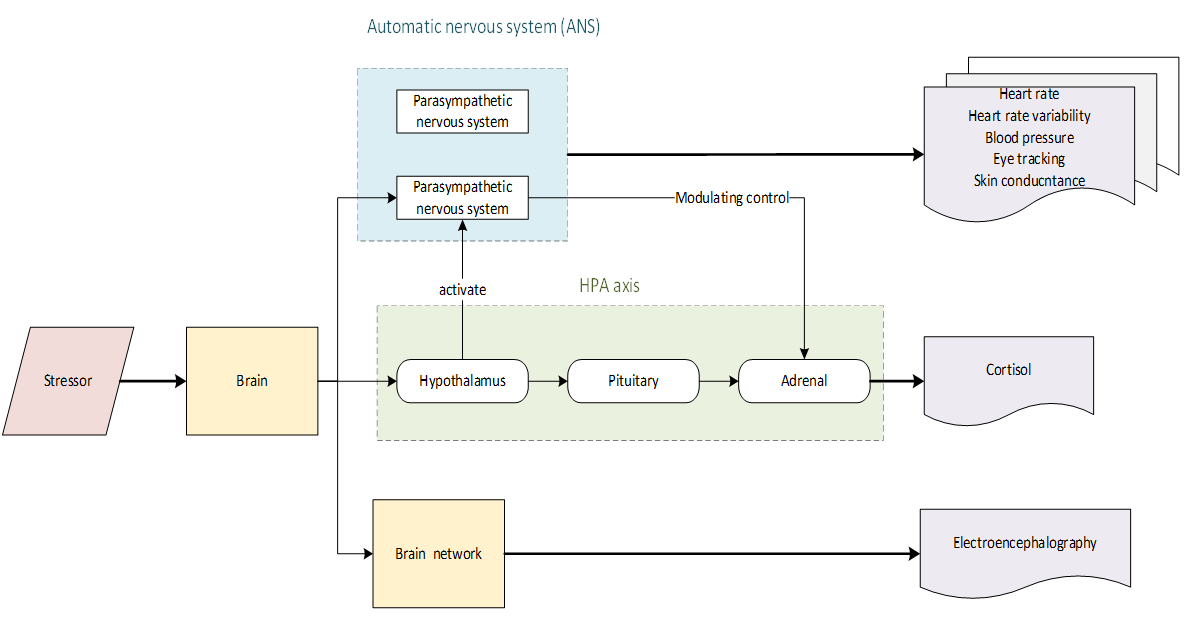}}
  \caption{\csentence{\label{fig:fg4}The physiological responses to stress}
      }
      \end{figure}

Several physiological responses to stress are involved during decision-making. A stressor will stimulate many organisms and various systems in the human body, including the automatic nervous system (ANS), the hypothalamic-pituitary-adrenal axis (HPA axis), and the brain network \cite{chrousos1992concepts}. Figure \ref{fig:fg4} illustrates the physiological responses to stress, which serves as the foundation for stress measurements.

ANS contains the sympathetic nervous system and the parasympathetic nervous system. Stressor stimulates the sympathetic nervous system and modulates control on the Adrenal medulla, which creates the synthesis and secretion of norepinephrine and epinephrine. The ANS triggers physiological responses, such as heart rate, HRV, blood pressure, eye movements, and skin conductance.

Stress can also induce an increased cortisol output via the HPA axis activation. When the stressor is applied to the HPA axis, the hypothalamus creates corticotropin-releasing factor (CRF), stimulating the pituitary. Then the pituitary secretes adrenocorticotropic hormone (ACTH), which stimulates the adrenal. The adrenal cortex is stimulated and secretes glucocorticoid (GC), which signals negative feedback to CRF and ACTH. Cortisol is the most critical human glucocorticoid, which is known as the stress hormone. It elevates blood sugar levels, enhances the brain’s use of glucose under stress conditions. 

Research over the past years has clarified that the brain network related to stress is involved in responding to stressors. With brain imaging technology development in mammals and the remarkable progress in genetic studies, the understanding of stress networks is rapidly growing in recent years. Stress networks are a set of highly connected brain structures that are activated when the animals perceive their surroundings or are exposed to various stressful events \cite{dallman2011regulation}. The brain network as shown in Figure \ref{fig:fg4} indicates EEG and brain region activation under stress.

Meanwhile, the body and nervous system’s organization and interactions reflect a high degree of complexity and multidirectional communication. Although stress involves many factors and various systems of the body, a stress reaction is a significant activation of stress response systems that result in a cascade of neuroendocrine changes \cite{dallman2011regulation}. Stress and related concepts, such as mental effort, workload, and cognitive workload, can be monitored and measured using physiological responses. Several physiological responses can be used as measurement metrics for stress, such as HRV, EEG, and NASA-TLX. Given the availability of multiple measurement methods, interests are increasing to identify the best methods and algorithms to measure and quantify stress. Researchers proposed a number of different measures. However, how to select the appropriate measures and metrics to measure and quantify stress remains a challenge.

Therefore, the literature review is based on the following four research postulates of stress in decision making: 
\begin{enumerate}[label=\arabic*)]
    \item  there is Bell-shaped relationship between stress and effort; 
    \item  stress is proportional to perceived workload, and inversely proportional to knowledge, skills, and affect; 
    \item there are two decision mechanisms: one is rational while the other is automatic. Both decision mechanism starts with a stressor as the input and ends with the output containing behaviours, stress, and cognitive workload. The output will recursively influence the decision process; and
    \item responses in the automatic nervous system (ANS), the hypothalamic-pituitary-adrenal axis (HPA axis), and the brain network can be used to measure stresses during a decision process.
\end{enumerate}

\subsection*{Protocol registration and reporting}
The systematic review will identify available measures for measuring stress, and evaluate and summarize the quantification algorithms and processing methods of different measures. This review will be conducted following the systematic review methodology and guideline by Normadhi, et al. \cite{normadhi2019identification}. The Preferred Reporting Items for Systematic Reviews and Meta-Analysis (PRISMA) framework will be used to guide the formulation of this systematic review protocol (see Additional file  1) \cite{moher2015preferred}. The completed PRISMA-P checklist is provided in Additional file 2.

\subsection*{Inclusion criteria}
\subsubsection*{Population}

We will identify studies investigating the effect of acute stress in disease-free adults. We will exclude experiments conducted among pediatrics participants or population with illness or pregnancies.

\subsubsection*{Outcomes}

The outcome of interest in this review is to explore the measurement, quantification, and application of emotional and rational decision process. This review focuses on studies investigating the theoretical or conceptual background of stress-related concepts. It also focuses on studies investigating the psychological measurement properties of stress-related concepts.
\subsubsection*{Type of studies}
We will select studies published in peer-reviewed scientific journals and related conferences. Studies need to present empirical data and assessed acute stress. The language restriction will be enforced on the condition that the English abstract is provided. Moreover, studies that reported bio-signal measures with sufficient quality will be included.  

\subsection*{Information sources and search strategy}

We will conduct comprehensive literature search in the following databases:

\begin{itemize}
\item Web of Science, 1990-2020
\item Scopus, 1990-2020
\item PubMed,1990-2020
\item ERIC (EBSCO), 1990-2020
\end{itemize}

Search terms related to stress, stress-related concepts, stress measurements, stress quantification, and stress application will be used. Searching uses a combination of keywords from different search terms. The search strategy is detailed in Additional file  3. We use the topic (including title, abstract, and keywords) search in the search design to retrieve the results.
This method is used to extract the most relevant research and reduce the number of articles that do not match the search. When the topic results are insufficient, other search structures (for example, title and abstract search) are performed.

\subsection*{Screening and selection procedure}

Two reviewers will screen titles, abstracts, and keywords independently. Conflicts will be
  resolved through discussion. A third reviewer will resolve any conflicts through consensus. The
  full text of all remaining articles will be obtained and assessed independently for inclusion by
  two reviewers. Discrepancies will be resolved by discussion, and any disagreement will be
 evaluated by a third reviewer. Articles selected for inclusion in the review must meet the predefined inclusion criteria.

\subsection*{Data extraction and synthesis}

For each included article, two reviewers will independently extract the data and verify the
  extracted data. If there is more than one publication reporting data from the same study, the
  first publication will be chosen to avoid double reporting data on the same participants. Two
  reviewers will also make a summary assessment of study validity and verify the assessment of
  validity. Any discrepancies will be resolved by discussion. 
  
  Data on the following factors will be extracted for each study: research objective, research
 method, research domain, triggering method, objective measure, subjective measure, and
 research results; the identification techniques and data analysis techniques. A standard
 information form to extract data from the selected articles \cite{normadhi2019identification}. Reviewers will use Mendeley to extract the essential information (e.g., title, authors, date of publication). Specific data will  be extracted from each article based on the initial study categorization. Excel spreadsheet will  be created and finalized after the two reviewers reviewed the data extraction. Reviewers will  check the consistency of the information extracted from each article to achieve high-quality  data extraction. Data extracted from the included articles will be synthesized separately for  stress measurement and quantification. The application, i.e., the research domain and research
 question, of stress will also be summarized.
 
 \subsection*{Quality assessment}
The methodological quality of each included study will be assessed using the Quality
 Assessment of Diagnostic Accuracy Studies-2 (QUADAS-2) tool \cite{whiting2011quadas}. The QUADAS-2 is used
 to assess the risk of bias and applicability concerns in the study. Each study will be assessed
 from 4 domains: participant/measure selection, index test, reference standard, and flow and
 timing \cite{kromm2021structure}. Classification of studies as low risk and high risk of bias and applicability will be  based on these four domains. Studies with missing information for any of the domains were  classified as having an unclear risk of bias or applicability for that specific domain. Other quality assessment tools will also be considered depending on the nature of final included studies.

\section*{Discussion}
Our systematic review aims to synthesize previous evidence on the definition of stress, and the
 measurement, quantification, and application of stress. We propose that human effort in
 decision-making is related to individual stress through a Bell-shaped curve. The decision
 mechanisms present that decision making is a dynamic process under stress. According to the
 stress level, the decision process can be classified into rational and automatic processes. We
 also aim to map the physiological responses to stress during the decision process to guide the
 identification of stress measures.

To the best of our knowledge, this will be the first systematic review to investigate measures
 for stress and all stress-related concepts. Performing an inclusive search in significant databases
 across a 30-year timescale will cover related research studies on this topic. We have provided
 a comprehensive overview of stress and stress-related concepts, decision mechanisms and the
physiological responses to stress in the protocol. Based on the relationship among stress and
 stress-related concepts, the decision mechanisms during rational and automatic decision
 processes are provided. 
 
 The systematic review will focus on investigating the measurement, quantification, and
 application of stress and stress-related concepts, including workload, cognitive workload, and
 mental effort. This review will aid technicians and researchers in selecting suitable stress
 measures for specific purpose. More importantly, appropriate stress measure selection will
 improve experiment objectivity, which will positive affect stress quantification. Clearly, there
 is a wide range of stressors and stress responses. We seek to identify the effective methods to
 measure and quantify stress and stress-related concepts, which will provide technical support
 and theoretical foundation for stress-related applications. Moreover, we will identify research
 trends from the current literature, which aims to establish a solid base for stress measurement
 and quantification and to guide further research. We expect our results to have significant 
implications in research and practice under stress such as ergonomics, human-computer interactions, design, learning and training, and so forth.

%%%%%%%%%%%%%%%%%%%%%%%%%%%%%%%%%%%%%%%%%%%%%%
%%                                          %%
%% Backmatter begins here                   %%
%%                                          %%
%%%%%%%%%%%%%%%%%%%%%%%%%%%%%%%%%%%%%%%%%%%%%%

\begin{backmatter}

\section*{Competing interests}
  The authors declare that they have no competing interests.
  
\section*{Abbreviations}
EEG: Electroencephalogram; HRV: Heart rate variability; NASA-TLX: NASA Task Load
 Index; DOI: Digital Object Identifier; URL: Uniform Resource Locator; ANS: Automatic
 nervous system; SNS: Sympathetic nervous system; PNS: Parasympathetic nervous system;
 HPA: Hypothalamic–pituitary–adrenal; CRF: Corticotrophin-releasing factor; ACTH:
 Adrenocorticotropic hormone; GC: Glucocorticoid; PRISMA-P: Preferred Reporting Items for
 Systematic Reviews and Meta-Analysis Protocol; QUADAS-2: Quality Assessment of
 Diagnostic Accuracy Studies-2.

\section*{Funding}
Funding was supported by NSERC Discovery Grant (RGPIN-2019-07048), NSERC CRD Project (CRDPJ 514052-17), and NSERC Design Chairs Program (CDEPJ 485989-14).

\section*{Author's contributions}
CS: manuscript writing. LY: critical revision of the manuscript for crucial conceptual
 improvement. XL: draft the manuscript and participate in early research. YZ: supervised
research, design and develop the research framework, critical revision of the manuscript. All
authors read and approved the final manuscript.

\section*{Author's information}
YZ has published substantially in design and stress. Specific topics of his research include the
 science of design, methodology for innovative and creative design, and neurocognitive model
 of design creativity. His academic program at Concordia University, Canada, focused on design
 and stress, which has led to the stress-effort model and its applications to multiple applications
 to design, learning, education, management, and health. CS is a doctoral student working on
 quantification of stress. LY is an epidemiologist, whose research program focuses on biological
 mechanisms between human movement and health, behavioural clinical trials, and
 implementation science. XL focuses on the neural signal processing.

\section*{Acknowledgements}
 Not applicable
%%  After submission of the .TEX file,                     %%
%%  you will be prompted to submit your .BBL file.         %%

% if your bibliography is in bibtex format, use those commands:
\bibliographystyle{bmc-mathphys} % Style BST file (bmc-mathphys, vancouver, spbasic).
\bibliography{bmc_article}      % Bibliography file (usually '*.bib' )
% for author-year bibliography (bmc-mathphys or spbasic)
% a) write to bib file (bmc-mathphys only)
% @settings{label, options="nameyear"}

% or include bibliography directly:
% \begin{thebibliography}
% \bibitem{b1}
% \end{thebibliography}

%%%%%%%%%%%%%%%%%%%%%%%%%%%%%%%%%%%
%%                               %%
%% Figures                       %%
%%                               %%
%% NB: this is for captions and  %%
%% Titles. All graphics must be  %%
%% submitted separately and NOT  %%
%% included in the Tex document  %%
%%                               %%
%%%%%%%%%%%%%%%%%%%%%%%%%%%%%%%%%%%

%%
%% Do not use \listoffigures as most will included as separate files

%%%%%%%%%%%%%%%%%%%%%%%%%%%%%%%%%%%
%%                               %%
%% Additional Files              %%
%%                               %%
%%%%%%%%%%%%%%%%%%%%%%%%%%%%%%%%%%%

\section*{Additional Files}
\subsection*{Additional file 1 --- PRISMA-P flow (DOCX 24 kb)}

\begin{figure}
\centering
\includegraphics[scale=0.6]{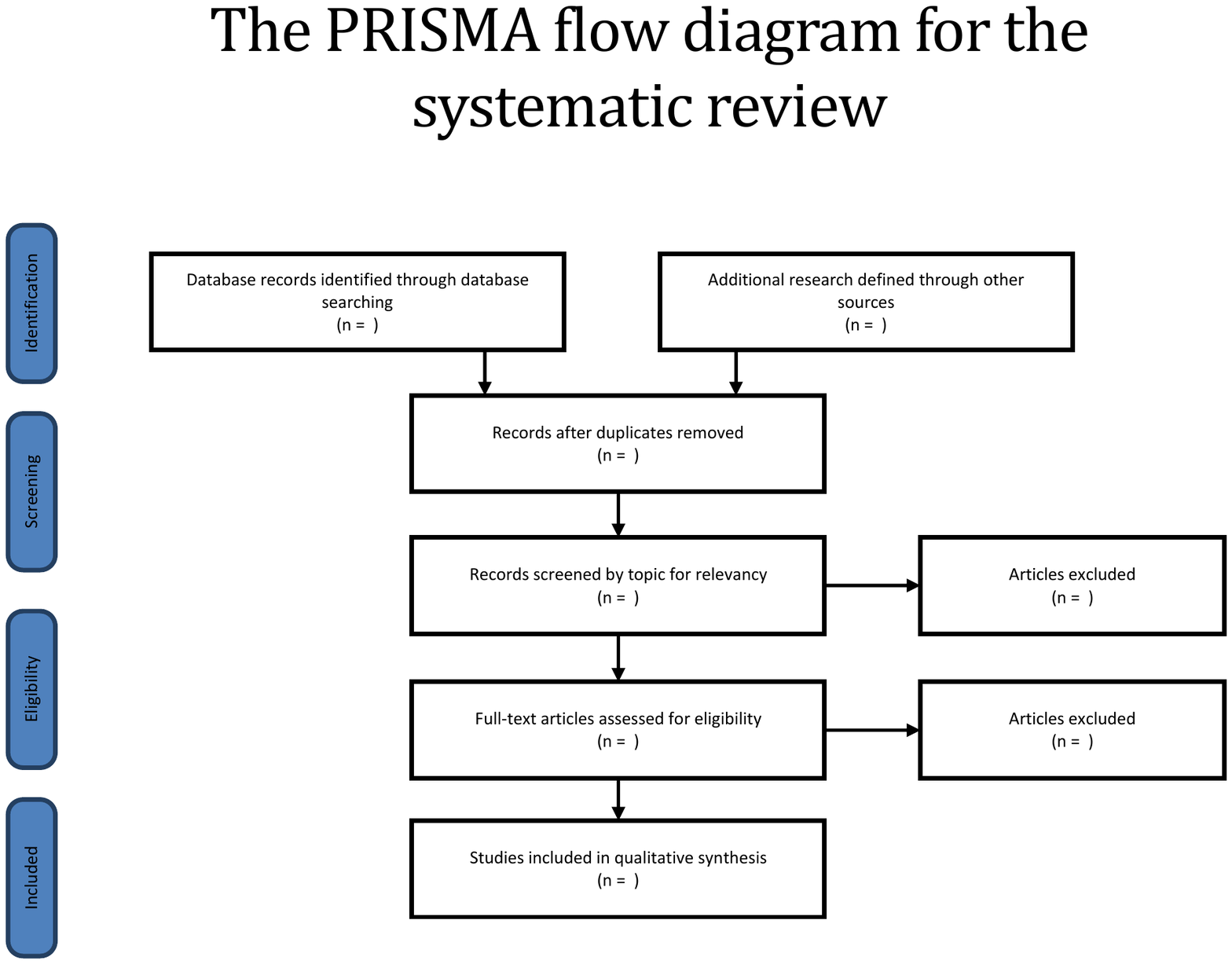}
\label{add1}
\end{figure}

\subsection*{Additional file 2 --- PRISMA-P checklist (DOCX 89 kb)}

\begin{figure}
\centering
\includegraphics[scale=0.6, angle = 90]{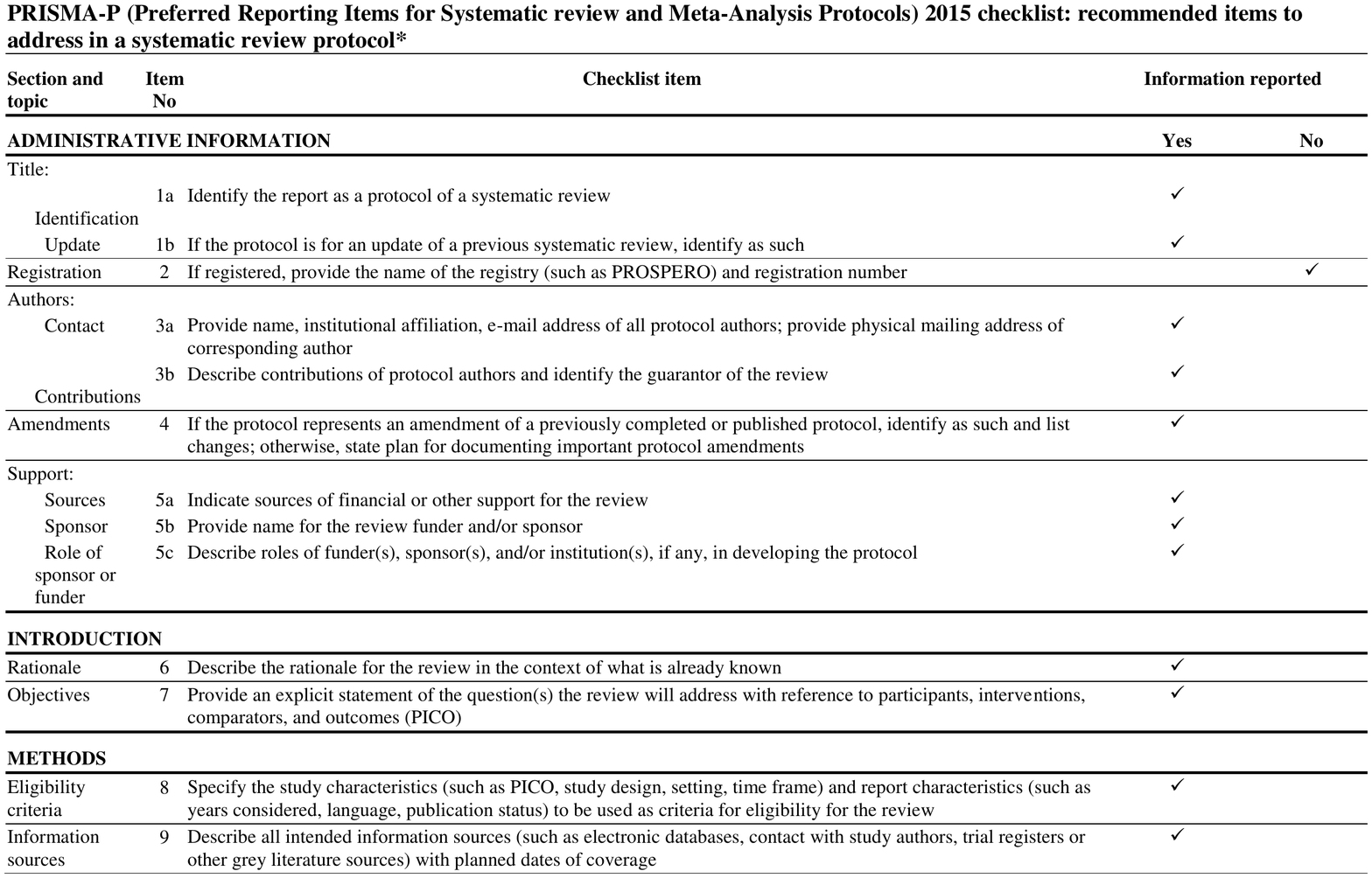}
\label{add2}
\end{figure}

\subsection*{Additional file 3 --- Search Strategy (DOCX 21 kb)}
\begin{figure}
\centering
\includegraphics[scale=0.6]{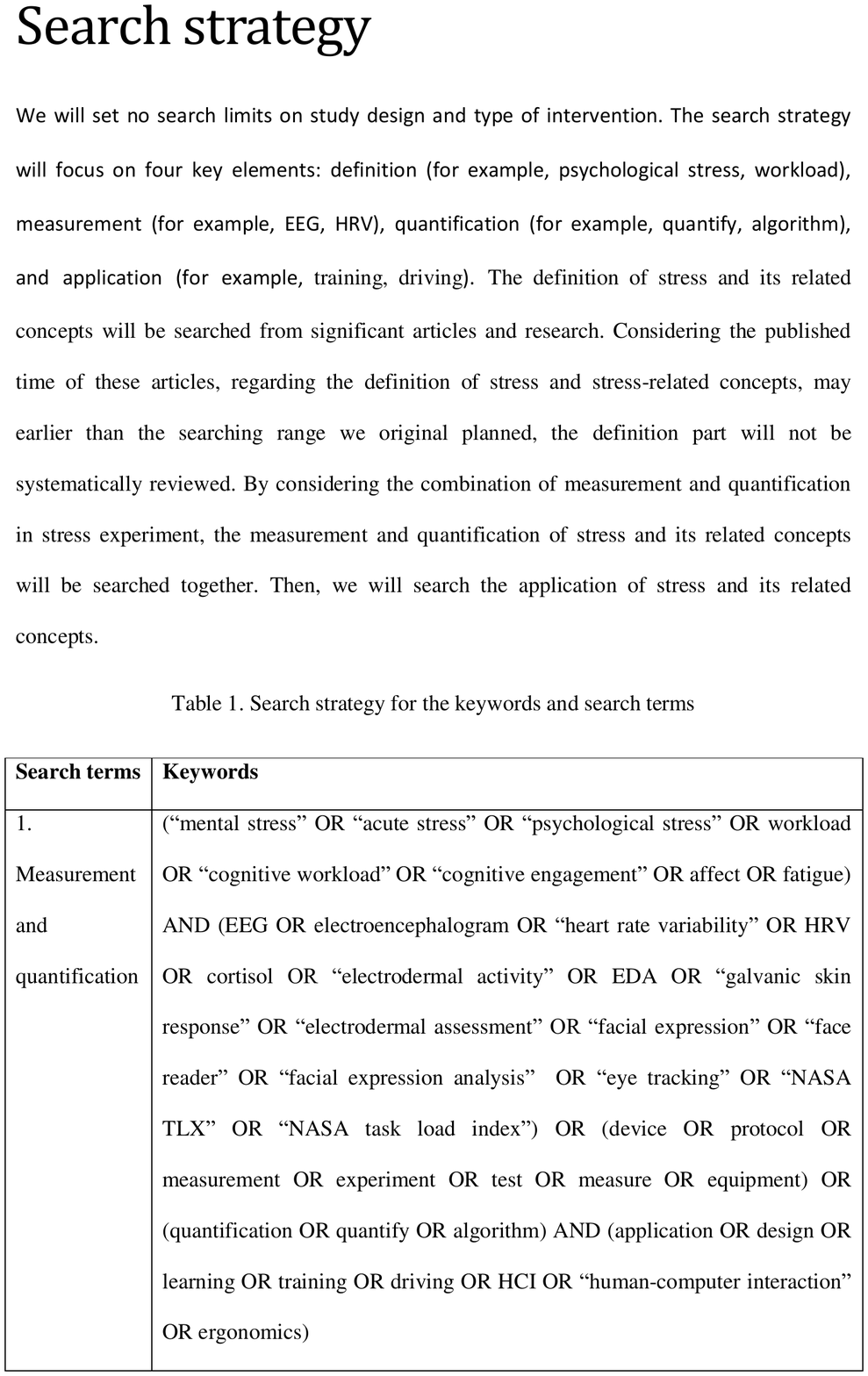}
\label{add3_1}
\end{figure}

\begin{figure}
\centering
\includegraphics[scale=0.6]{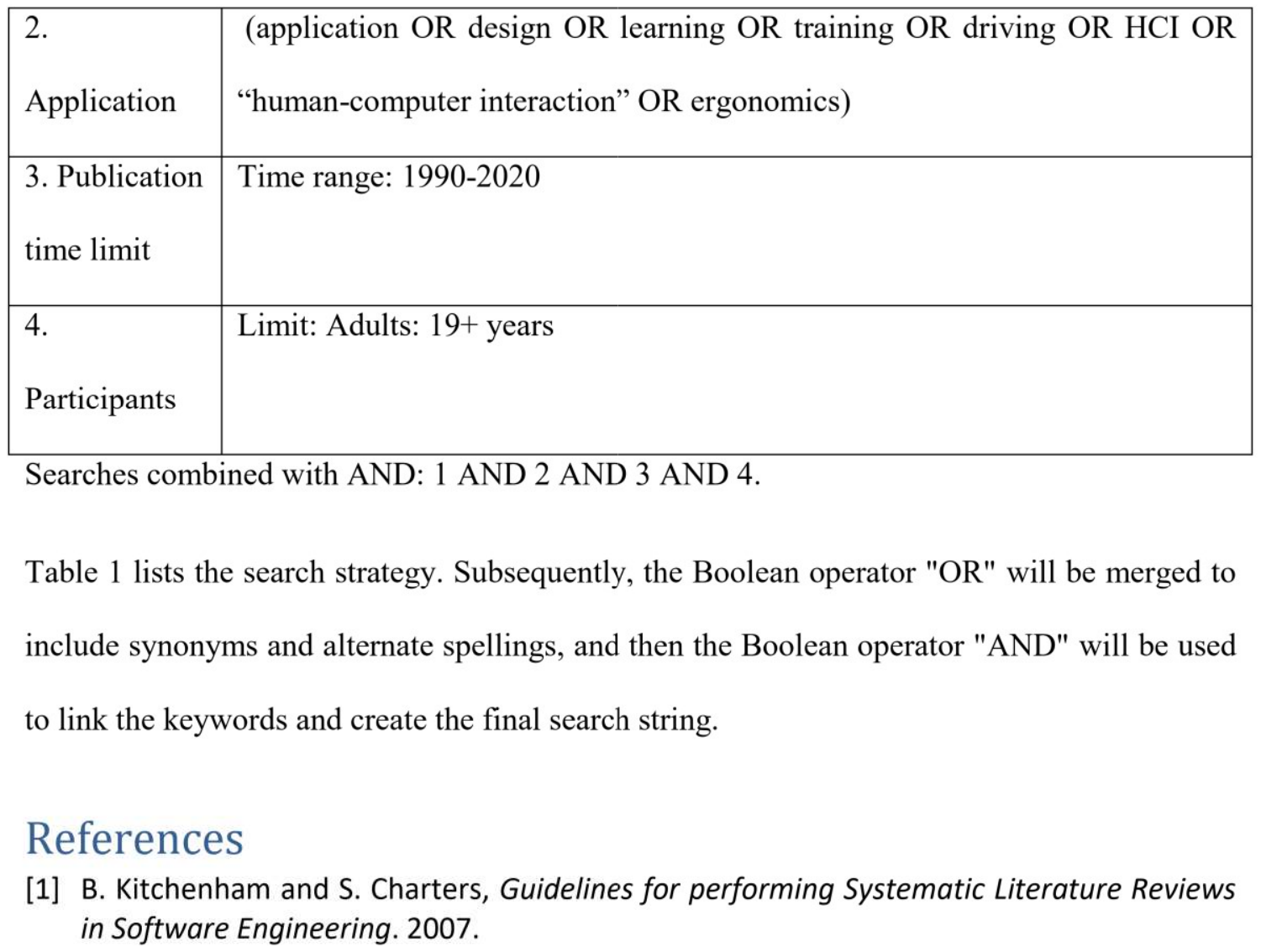}
\label{add3_2}
\end{figure}

\end{backmatter}
\end{document}